\documentclass[twocolumn,prl,floats,showpacs]{revtex4} 
\usepackage{amsfonts}
\usepackage{amsmath}
\usepackage{amssymb}
\usepackage{graphicx} 
\begin{document}
\title{Quantum atom optics with fermions from molecular dissociation}
\author{K. V. Kheruntsyan}
\affiliation{ARC Centre of Excellence for Quantum-Atom Optics, School of Physical Sciences,
University of Queensland, Brisbane, Qld 4072, Australia}
\date{\today {}}

\begin{abstract}
We study a fermionic atom optics counterpart of parametric down-conversion
with photons. This can be realized through dissociation of a Bose-Einstein
condensate of molecular dimers consisting of fermionic atoms. We present a
theoretical model describing the quantum dynamics of dissociation and find
analytic solutions for mode occupancies and atomic pair correlations, valid in
the short time limit. The solutions are used to identify upper bounds for the
correlation functions, which are applicable to any fermionic system and
correspond to ideal particle number-difference squeezing.

\end{abstract}

\pacs{03.75.-b, 03.65.-w, 05.30.-d, 42.50.-p}
\maketitle

Advances in the experimental control of degenerate quantum gases of neutral
atoms have recently reached the stage where atomic correlations and quantum
statistics can be \textit{directly} accessed via the measurement of atom shot
noise and atom counting \cite{Greiner,Bloch,Raizen,Esslinger,Aspect}. Given
the similarities with the pioneering photon correlation measurements of
Hanbury Brown and Twiss \cite{HBT} and the intriguing parallels with modern
quantum optics, these experiments represent a remarkable step forward in
advancing quantum atom optics. Earlier and related experiments were performed
either in a cold but not degenerate atomic beam \cite{Yasuda-Shimizu} or else
were examples of \textit{indirect} measurements of higher-order correlations
and number squeezing \cite{Phillips-1997,Kasevich,Tolra,Weiss}.

In quantum optics with photons the most successful applications have been
achieved using squeezed light and entangled photon pairs from parametric down
conversion \cite{Kimble,Heidman,JOSA-B-1987,Walls}. A matter-wave or atom
optics counterpart of down conversion can be realized through dissociation of
a Bose-Einstein condensate (BEC) of molecular dimers. The most direct analogy
with quantum optics corresponds to the case of bosonic statistics of
constituent atom pairs \cite{Moelmer2001,TwinBeams,Yurovsky,EPR}, as realized
in dissociation experiments with $^{23}$Na$_{2}$ and $^{87}$Rb$_{2}$
\cite{Dissociation-exp-Ketterle,Dissociation-exp-Rempe}. In contrast to this,
the recent correlation measurements at JILA \cite{Greiner}, using dissociation
of $^{40}$K$_{2}$ molecules near a magnetic Feshbach resonance, have an
intriguing twist in that the constituent atoms obey fermionic statistics. In
this case, analogies with photonic down conversion and the implications for
possible future applications are not so immediate as in the case of bosons.

In this paper we study the quantum dynamics of dissociation of diatomic
molecules into fermionic atoms and analyze the resulting atom correlations in
momentum space. At low densities and small $s$-wave interactions the
momentum correlations are reflected in the density correlations after spatial
expansion of the cloud. Accordingly, the results obtained here are related to
the spatial correlation measurements performed at JILA \cite{Greiner}.
Together with the recent studies of fermionic four-wave mixing
\cite{FWM-fermions} and association of fermionic atoms into molecules
\cite{Javanainen,Meystre-association}, the present work 
(see also \cite{Jack-Pu}) expands the paradigm
of \emph{fermionic quantum atom optics}. Possible applications are in precision
measurements and fundamental tests of quantum mechanics, similar to those
proposed recently for dissociation into bosonic atoms
\cite{TwinBeams,EPR}.

Due to the limitations of the undepleted molecular field approximation
employed here, the obtained results are only applicable to short dissociation
times. However, the advantage is the analytic transparency of the results,
which provide useful insights at the conceptual level. In particular, we show
rigorously that the obtained pair correlations represent their generic upper
bounds and are applicable to any fermion system. This gives a useful reference
for further (numerical) studies of this and related systems with less
restricted approximations. In addition, we point out that the notions of
maximum correlation and squeezing of atom number-difference fluctuations have
to take into account the fact that the shot noise level for fermions is
fundamentally different to what one usually encounters in quantum optics with bosons.

We start the analysis by considering an effective field theory Hamiltonian for
the coupled atomic-molecular system given by $\hat{H}=\hat{H}_{0}-$
$i\hbar\chi\int d\mathbf{x(}\hat{\Psi}_{0}^{\dagger}\hat{\Psi}_{\uparrow}^{{} 
}\hat{\Psi}_{\downarrow}^{{}}-\hat{\Psi}_{\downarrow}^{\dagger}\hat{\Psi
}_{\uparrow}^{\dagger}\hat{\Psi}_{0}^{{}})$ \cite{PDKKHH}. Here, $\hat{H}_{0}$
stands for the usual kinetic energy term plus the trapping potential,
$\hat{\Psi}_{0}^{{}}(\mathbf{x},t)$ is the bosonic field operator for
molecules of mass $m_{0}$, while $\hat{\Psi}_{\uparrow(\downarrow)} 
(\mathbf{x},t)$ are fermionic operators for atoms (with masses $m_{\uparrow
(\downarrow)}=m_{0}/2\equiv m$) in two different spin states, $\sigma
=\uparrow,\downarrow$. The atom-molecule coupling is described by $\chi$, and
we have omitted intra- and inter-species $s$-wave scattering interaction
terms, which is justified at low particle densities.

Considering a uniform system in a cubic box of side $L$, we employ an
undepleted molecular field approximation valid in the short time limit. The
molecular field is described by a coherent state and we absorb its mean-field
amplitude $\Psi_{0}$ into an effective coupling $g=\chi\sqrt{n_{0}}$, where
$n_{0}=|\Psi_{0}|^{2}$ is the density. Assuming periodic boundary conditions
and expanding the atomic fields in a plane-wave basis, in terms of single-mode
operators, we obtain the following effective Hamiltonian, in a rotating frame: 
\begin{equation}
\hat{H}=\sum\nolimits_{\mathbf{k},\sigma}\hbar\Delta_{\mathbf{k}}\hat
{n}_{\mathbf{k}\sigma}-i\hbar g\sum\nolimits_{\mathbf{k}}(\hat{c} 
_{\mathbf{k}\uparrow}^{{}}\hat{c}_{-\mathbf{k}\downarrow}^{{}}-\hat
{c}_{-\mathbf{k}\downarrow}^{\dagger}\hat{c}_{\mathbf{k}\uparrow}^{\dagger}).
\label{H2} 
\end{equation}
Here, $\hat{c}_{\mathbf{k}\sigma}^{\dagger}$ ($\hat{c}_{\mathbf{k}\sigma}$)
are fermionic creation (annihilation) operators, $\hat{n}_{\mathbf{k}\sigma
}=\hat{c}_{\mathbf{k}\sigma}^{\dagger}\hat{c}_{\mathbf{k\sigma}}$ is the
particle number operator,$\ \mathbf{k}$ is the momentum [$k_{i}=2\pi n_{i}/L$,
$n_{i}=0,\pm1,\pm2,\ldots,$ $i=x,y,z$] and $\Delta_{\mathbf{k}}\equiv
\Delta+\hbar\mathbf{k}^{2}/(2m)$. The detuning $2\Delta$ corresponds to the
overall energy mismatch $2\hbar\Delta$ between the free two-atom state in the
dissociation threshold and the bound molecular state.

The model system in mind corresponds to a pure molecular BEC on the stable
side of a Feshbach resonance, with no residual atoms present. This is followed
by a rapid switching on of the coupling $\chi$ (e.g., via an rf transition or
a rapid crossing through the resonance to the atomic side) and a simultaneous
switching off of the trapping potential. From this stage onward, the atomic
filed evolves in free space, with a vacuum initial state and negative detuning
$\Delta$. Since the excess of energy $2\hbar|\Delta|$ is released into the
kinetic energy of dissociated atom pairs in the two spin states,
$2\hbar|\Delta|\rightarrow\hbar^{2}(|\mathbf{k}_{\uparrow}|^{2}+|\mathbf{k} 
_{\downarrow}|^{2})/(2m)$, we expect -- from momentum conservation -- strong
correlation between the atoms with opposite spins and momenta, $\mathbf{k} 
_{\uparrow}=-\mathbf{k}_{\downarrow}$. In fact, the interaction term in the
Hamiltonian (\ref{H2}) is the prototype interaction to produce -- in the
lowest order perturbation theory -- an entangled spin singlet state. It is
also a fermionic analog of the squeezing Hamiltonian in quantum optics
\cite{Walls}.

Introducing a dimensionless time $\tau=t/t_{0}$, length $l=L/d_{0}$, detuning
$\delta=\Delta t_{0}$, and momentum $\mathbf{q}=\mathbf{k}d_{0}$, where
$t_{0}=1/g$ is the time scale and $d_{0}=\sqrt{\hbar t_{0}/(2m)}$ is the
length scale, we can put the system into a dimensionless form, with the
Heisenberg equations of motion
\begin{subequations}
\label{Heis} 
\begin{align}
d\hat{c}_{\mathbf{q}\uparrow}/d\tau &  =-i\delta_{\mathbf{q}}\hat
{c}_{\mathbf{q}\uparrow}-\hat{c}_{-\mathbf{q}\downarrow}^{\dagger
},\label{Heis-a}\\
d\hat{c}_{-\mathbf{q}\downarrow}^{\dagger}/d\tau &  =i\delta_{\mathbf{q}} 
\hat{c}_{-\mathbf{q}\downarrow}^{\dagger}+\hat{c}_{\mathbf{q}\uparrow},
\label{Heis-b} 
\end{align}
where $\delta_{\mathbf{q}}\equiv q^{2}+\delta$ ($q=|\mathbf{q}|$). Solutions
to Eqs. (\ref{Heis}) are:
\end{subequations}
\begin{subequations}
\label{sol} 
\begin{align}
\hat{c}_{\mathbf{q}\uparrow}(\tau)  &  =A_{\mathbf{q}}(\tau)\hat
{c}_{\mathbf{q}\uparrow}(0)-B_{\mathbf{q}}(\tau)\hat{c}_{-\mathbf{q} 
\downarrow}^{\dagger}(0),\label{sol-a}\\
\hat{c}_{-\mathbf{q}\downarrow}^{\dagger}(\tau)  &  =B_{\mathbf{q}}(\tau
)\hat{c}_{\mathbf{q}\uparrow}(0)+A_{\mathbf{q}}^{\ast}(\tau)\hat
{c}_{-\mathbf{q}\downarrow}^{\dagger}(0). \label{sol-b} 
\end{align}
Here, $A_{\mathbf{q}}(\tau)=\cos\left(  g_{\mathbf{q}}\tau\right)
-i\delta_{\mathbf{q}}\sin\left(  g_{\mathbf{q}}\tau\right)  /g_{\mathbf{q}}$,
$B_{\mathbf{q}}(\tau)=\sin\left(  g_{\mathbf{q}}\tau\right)  /g_{\mathbf{q}}$,
$g_{\mathbf{q}}\equiv(1+\delta_{\mathbf{q}}^{2})^{1/2}$, and $|A_{\mathbf{q} 
}|^{2}+B_{\mathbf{q}}^{2}=1$. Using these solutions with vacuum initial
conditions, we find that the only nonzero second-order moments are the mode
occupancies and the pairing fields:
\end{subequations}
\begin{align}
n_{\mathbf{q}}(\tau)  &  \equiv\left\langle \hat{n}_{\mathbf{q}\sigma} 
(\tau)\right\rangle =B_{\mathbf{q}}^{2}(\tau)=\sin^{2}\left(  g_{\mathbf{q} 
}\tau\right)  /g_{\mathbf{q}}^{2},\label{n}\\
m_{\mathbf{q}}(\tau)  &  \equiv\left\langle \hat{c}_{\mathbf{q}\uparrow} 
(\tau)\hat{c}_{-\mathbf{q}\downarrow}(\tau)\right\rangle =A_{\mathbf{q}} 
(\tau)B_{\mathbf{q}}(\tau), \label{m} 
\end{align}
which, in addition, are related by
\begin{equation}
|m_{\mathbf{q}}(\tau)|^{2}=[1-n_{\mathbf{q}}(\tau)]n_{\mathbf{q}}(\tau).
\label{m-amplitude} 
\end{equation}

For comparison, in the case of bosons the last terms in Eqs.~(\ref{Heis-a})
and (\ref{sol-a}) acquire positive signs, while the $\sin$ ($\cos$) terms in
the coefficients $A_{\mathbf{q}}$ and $B_{\mathbf{q}}$ are replaced by $\sinh$
($\cosh$), together with $g_{\mathbf{q}}\equiv(1-\delta_{\mathbf{q}} 
^{2})^{1/2}$ and $|A_{\mathbf{q}}|^{2}-B_{\mathbf{q}}^{2}=1$ \cite{Comment}.
As a result, the solutions give $n_{\mathbf{q}}(\tau)=\sinh^{2}\left(
g_{\mathbf{q}}\tau\right)  /g_{\mathbf{q}}^{2}$, and $|m_{\mathbf{q}} 
(\tau)|^{2}=[1+n_{\mathbf{q}}(\tau)]n_{\mathbf{q}}(\tau)$.

\begin{figure}[ptb]
\includegraphics[height=3.45cm]{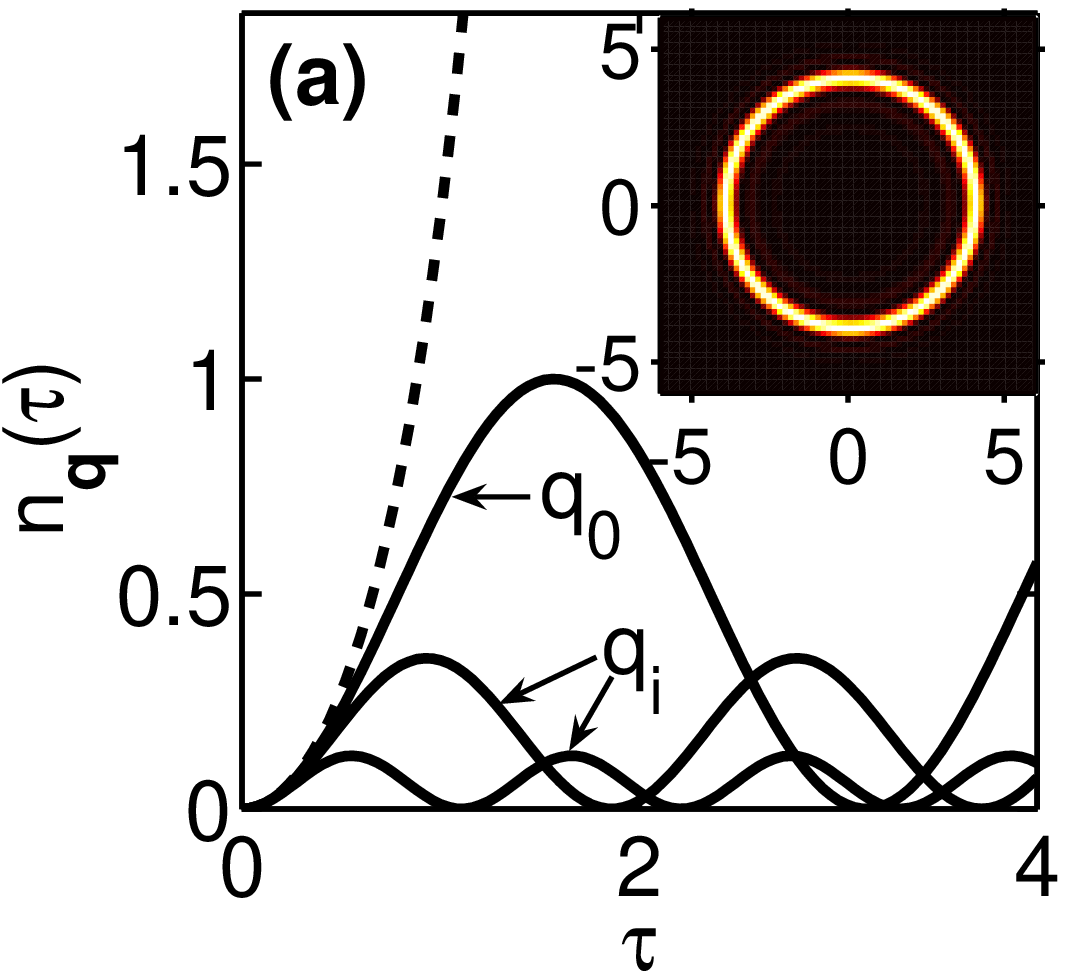}~~~~\includegraphics[height=3.45cm]{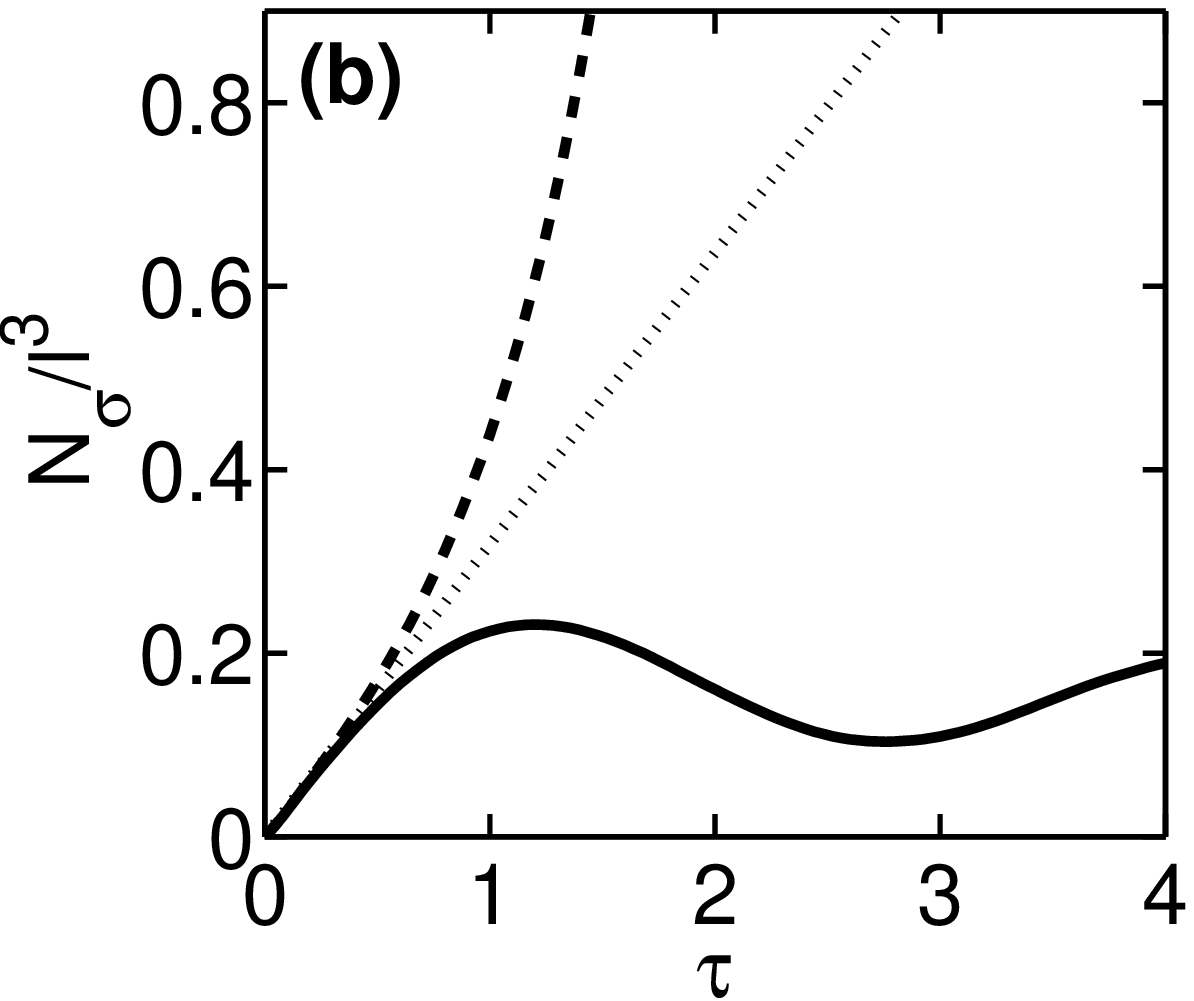}
\caption{(a) Average occupancies of one of the resonant modes $q_{0}=$
$\sqrt{|\delta|}$ and two sidebands $q_{i}$ (full lines) as a function of
time, for $\delta=-16$. The dashed line is the solution for the $q_{0}$-mode
in the case of bosons, which grows exponentially. The inset shows the slice
through the origin of the fermionic 3D momentum distribution $n_{\mathbf{q} 
}(\tau)$ at $\tau=0.6$. (b) Total number of atoms $N_{\sigma}/l^{3}$ (full
line) as a function of time, for $\delta=-16$. The normalization with respect
to $l^{3}$ makes this quantity independent of the quantization volume. The
dashed line is the respective bosonic result, while the straight dotted line
is the total number of atoms $N_{\sigma}(\tau)/l^{3}\simeq\sqrt{|\delta|} 
\tau/(4\pi)$ obtained using Fermi's golden rule calculation of the molecular
spontaneous decay rate \cite{Dissociation-exp-Ketterle} in the linear regime.} 
\label{fig1} 
\end{figure}

From Eq.~(\ref{n}) we see that the momentum distribution of the atoms in the
two spin states, $\uparrow$ and $\downarrow$, is the same. The average mode
occupancies undergo oscillations characteristic of fermionic statistics (see
Fig. \ref{fig1}a); the maximum occupancy of $n_{\mathbf{q}}(\tau)=1$ imposed
by the Pauli exclusion principle is reached at integer multiples of time
$\tau=\pi/2$, for resonant modes satisfying $g_{\mathbf{q}_{0}}=1$. For
$\mathbf{q}_{0}$ to be nonzero, this condition requires a negative detuning
$\delta$, and therefore the absolute resonant momentum is given by
$q_{0}=|\mathbf{q}_{0}|=\sqrt{|\delta|}$. During the initial stage
($\tau\lesssim0.6$), the occupancies grow in-phase and the 3D momentum
distribution is peaked on the surface of a spherical shell of radius $q_{0}$
as shown in the inset of Fig. \ref{fig1}a. At later times the oscillations
dephase and the distribution function becomes more complicated in structure.

The total number of atoms in each spin state, $N_{\sigma}(\tau)=\sum
\nolimits_{\mathbf{q}}n_{\mathbf{q}}(\tau)$, as a function of time is shown in
Fig. \ref{fig1}b. The initial growth of $N_{\sigma}(\tau)$ saturates at
$\tau\simeq1.2$, after which we see non-trivial oscillations. This is a
combined effect of Pauli blocking and the oscillatory behavior of the
individual mode occupancies. We emphasize that the saturation in this model is
obtained within the undepleted molecular field approximation, and thus is
purely a consequence of Fermi statistics. By comparison, the same
approximation for bosons leads to an exponentially growing output due to
bosonic stimulation and hence to unphysical results in the long time limit.
Here, once the depletion is taken into account, the saturation is naturally
reached due to a finite initial number of molecules.

In this respect, the present fermionic results suggest that the undepleted
molecular field approximation for fermions is more reliable than for bosons.
At the level of pairwise mode coupling, this conjecture is in fact supported
by the results of a numerical simulation using an exact quantum Monte Carlo
method \cite{Corney-Drummond}. For example, at $\tau\simeq\pi/2$ the
discrepancy between the present and the exact results is about $5\%$ for
fermions, while it is $\sim11\%$ for bosons and grows further with time.
Similarly, the negligible role of the molecular depletion in the short time
limit can be verified using the results of a related numerical study of Ref.
\cite{Jack-Pu} where the molecular field dynamics is treated at the mean-field level.

\begin{figure}[ptb]
\includegraphics[height=3.78cm]{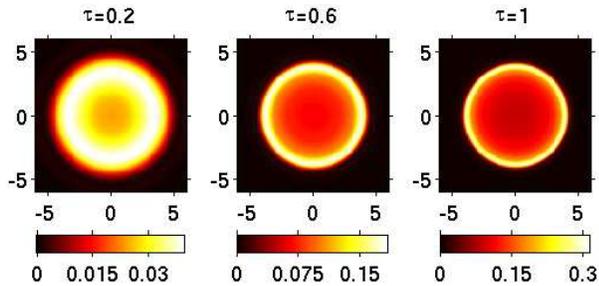} \caption{Snapshots of the average
column density in momentum space $\overline{n}_{\mathbf{p}}(\tau)/l$ at
different times $\tau$, for $\delta=-16$.} 
\label{fig2} 
\end{figure}

We now turn to the analysis of pair correlations of the atoms in the opposite
spin states and consider a normalized correlation function between 3D density
fluctuations $\Delta\hat{n}_{\mathbf{q}\sigma}=\hat{n}_{\mathbf{q}\sigma
}-\langle\hat{n}_{\mathbf{q}\sigma}\rangle$ in momentum space:
\begin{equation}
g_{\uparrow\downarrow}(\mathbf{q},\mathbf{q}^{\prime},\tau)=\langle\Delta
\hat{n}_{\mathbf{q}\uparrow}\Delta\hat{n}_{\mathbf{q}^{\prime}\downarrow
}\rangle/\sqrt{\langle\hat{n}_{\mathbf{q}\uparrow}\rangle\langle\hat
{n}_{\mathbf{q}^{\prime}\downarrow}\rangle}. \label{corr} 
\end{equation}

For atom pairs with non-opposite momenta the pair correlation vanishes,
$\left.  g_{\uparrow\downarrow}(\mathbf{q},\mathbf{q}^{\prime},\tau
)\right\vert _{\mathbf{q}^{\prime}\neq-\mathbf{q}}=0$, implying the absence of
any correlation. In the case of equal but opposite momenta, we find that
\begin{equation}
g_{\uparrow\downarrow}(\mathbf{q},-\mathbf{q},\tau)=|m_{\mathbf{q}}(\tau
)|^{2}/n_{\mathbf{q}}(\tau)=1-n_{\mathbf{q}}(\tau)<1. \label{g-diss} 
\end{equation}
This corresponds to the maximum degree of correlation -- as a consequence of
Eq.\thinspace(\ref{m-amplitude}), except when $n_{\mathbf{q}}(\tau)=1$ in
which case $g_{\uparrow\downarrow}(\mathbf{q},-\mathbf{q},\tau)$ coincides
with the uncorrelated level. For bosons, the respective pair correlation is
given by $g_{\uparrow\downarrow}(\mathbf{q},-\mathbf{q},\tau)=1+n_{\mathbf{q} 
}(\tau)$, which increases with $n_{\mathbf{q}}(\tau)$ and always stays above zero.

In order to make a better connection with the experiments at JILA
\cite{Greiner} we note that the correlation measurements were made using
absorption images after a time-of-flight expansion. This corresponds to
analyzing the spatial column densities which involve integration of the 3D
density along the direction of propagation of the imaging laser. Accordingly,
we now analyze the momentum space analog of this procedure and calculate the
correlation between momentum column density fluctuations.

\begin{figure}[ptb]
\includegraphics[height=3.4cm]{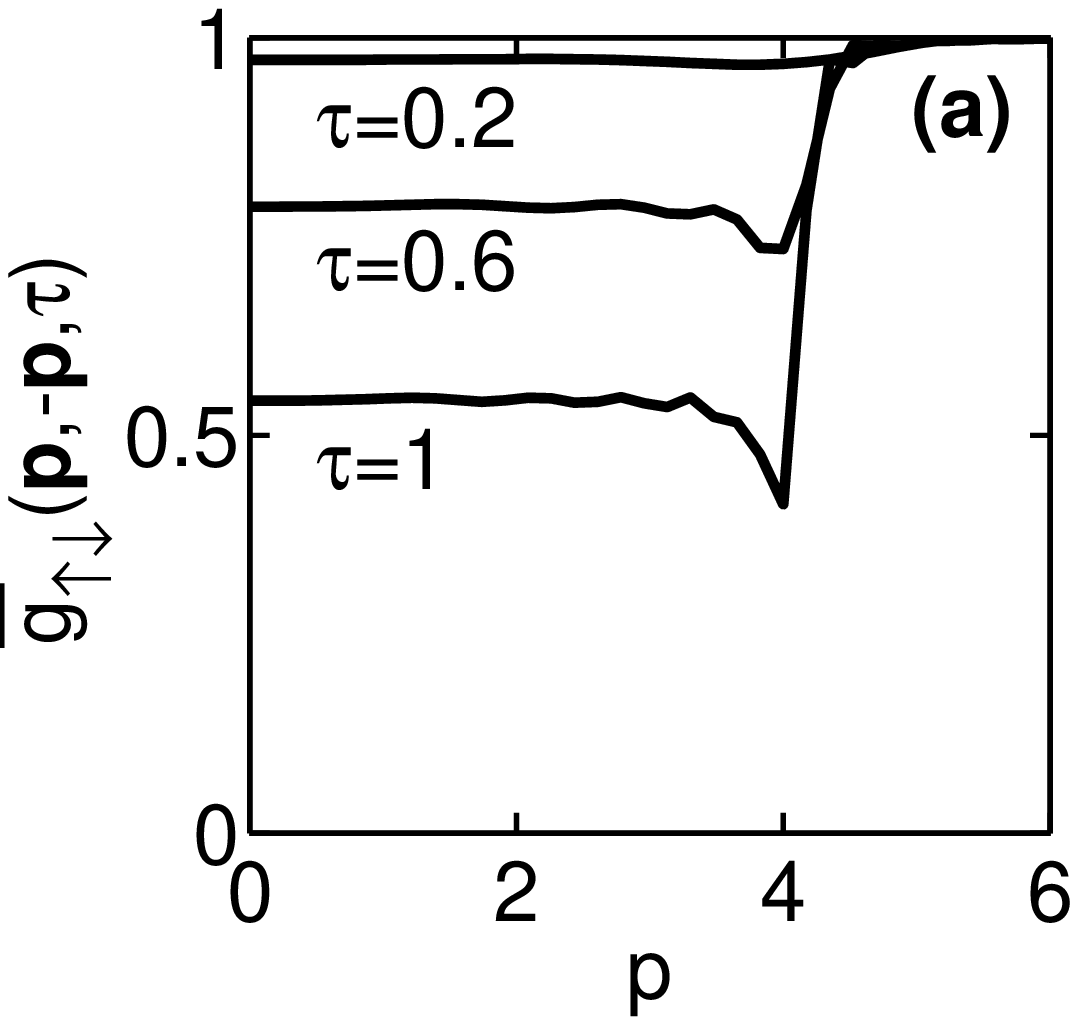}~~~~\includegraphics[height=3.4cm]{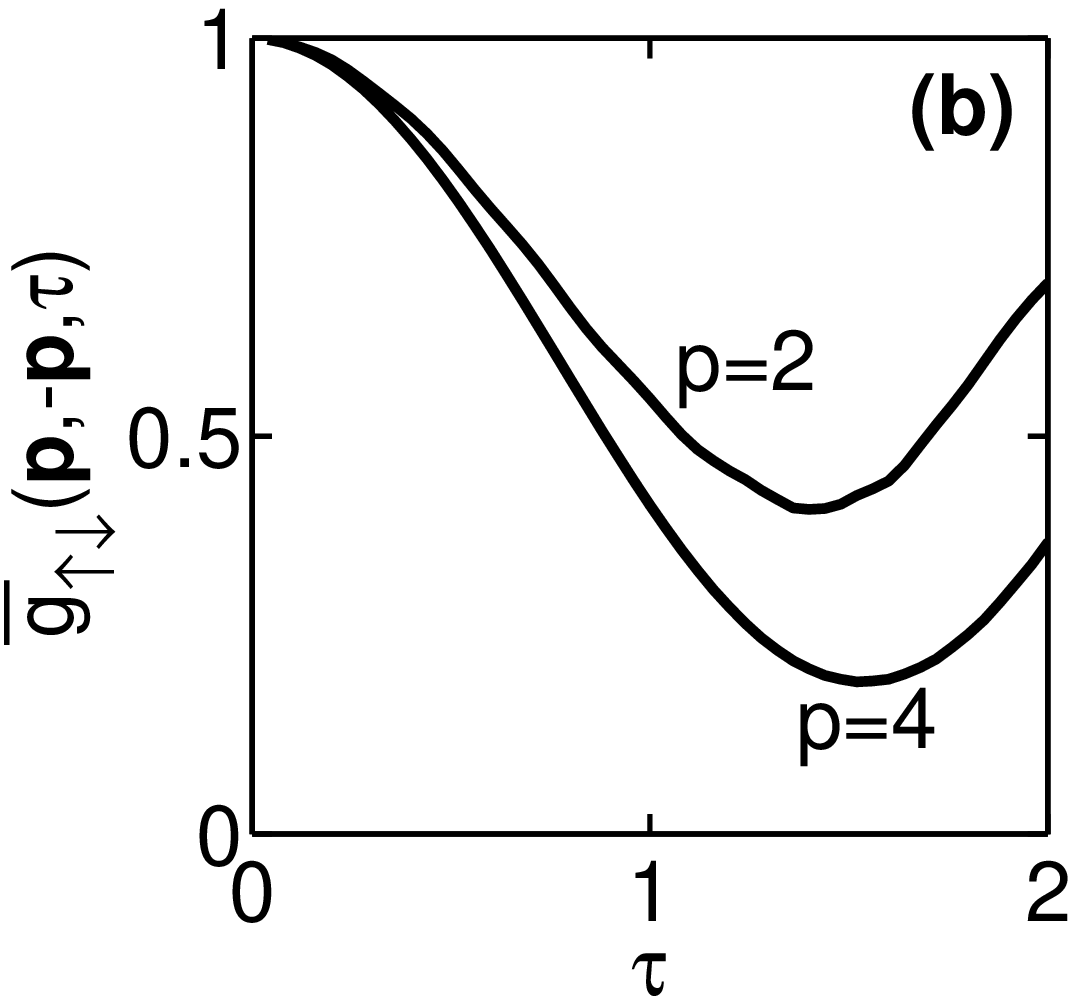}
\caption{(a) Correlation between momentum column density fluctuations,
$\overline{g}_{\uparrow\downarrow}(\mathbf{p},-\mathbf{p},\tau)$, as a
function of the absolute 2D momentum $p=|\mathbf{p}|$ at different times
$\tau$, for $\delta=-16$. The correlation is larger initially, in the
few-particle regime, and decreases as the number of atoms grows. Similar
relationship is seen within the low-density tails and the higher-density
central part of the momentum distribution. (b) Correlation signal as a
function of time, for two different values of the $p$.} 
\label{fig3} 
\end{figure}

The atom number operator corresponding to a $z$-integrated momentum column
density is given by $\widehat{\overline{n}}_{\mathbf{p}\sigma}=\sum
\nolimits_{q_{z}}\hat{n}_{\mathbf{q}\sigma}$, where $\mathbf{p\equiv} 
(q_{x},q_{y})$ is the reduced 2D momentum. Using Eq. (\ref{n}), the average
column density is found via $\overline{n}_{\mathbf{p}}(\tau)=\langle
\widehat{\overline{n}}_{\mathbf{p}\sigma}(\tau)\rangle=\sum\nolimits_{q_{z} 
}n_{\mathbf{q}}(\tau)$ and is the same for the two spin states. Snapshots of
$\overline{n}_{\mathbf{p}}(\tau)$ at different times $\tau$ are plotted in
Fig. \ref{fig2}. The last two frames show a clear ring structure around the
central background, which is consistent with the observed absorption images
\cite{Greiner} of spatial column densities after free expansion.

The correlation function between momentum column density fluctuations in the
two spin states, which we denote via $\overline{g}_{\uparrow\downarrow
}(\mathbf{p},\mathbf{p}^{\prime},\tau)$, is defined as in Eq. (\ref{corr})
except that the operators $\hat{n}_{\mathbf{q}\sigma}$ are replaced by
$\widehat{\overline{n}}_{\mathbf{p}\sigma}$. In this case, the bars above the
operators signify the procedure of summation over the $q_{z}$-component,
before taking the ensemble average. For equal but opposite momenta,
$\mathbf{p}^{\prime}=-\mathbf{p}$, we find (see Fig. \ref{fig3}) that
\begin{equation}
\overline{g}_{\uparrow\downarrow}(\mathbf{p},-\mathbf{p},\tau)=1-\sum
\nolimits_{q_{z}}\left[  n_{\mathbf{q}}(\tau)\right]  ^{2}\left/  \overline
{n}_{\mathbf{p}}(\tau)\right.  <1. \label{corr-CD} 
\end{equation}
For any other pair of momenta ($\mathbf{p}^{\prime}\neq-\mathbf{p}$), the
correlation function is simply zero, implying the absence of any correlation.
In the case of dissociation into bosonic atoms, the same correlation function
is given by $\overline{g}_{\uparrow\downarrow}(\mathbf{p},-\mathbf{p} 
,\tau)=1+\sum\nolimits_{q_{z}}\left[  n_{\mathbf{q}}(\tau)\right]  ^{2}\left/
\overline{n}_{\mathbf{p}}(\tau)\right.  >1$.

It is important to point out that the degree of correlation between atom pairs
with opposite spins and momenta obtained in this model is maximal at any given
density. The notion of maximal is defined here to correspond to perfect
(100\%) noise reduction of the number-difference fluctuations below the
shot-noise level.

This can be easily understood at the level of just two modes $\hat
{c}_{\mathbf{q\uparrow}}$ and $\hat{c}_{-\mathbf{q\downarrow}}$, which we
define via $\hat{c}_{1}$ and $\hat{c}_{2}$. Considering the normalized
variance of the particle number-difference fluctuations, $V=\left\langle
[\Delta(\hat{n}_{1}-\hat{n}_{2})]^{2}\right\rangle /SN$, where $SN$ is the
shot noise level, one can show that in the simplest case of $\left\langle
\hat{n}_{1}\right\rangle =\left\langle \hat{n}_{2}\right\rangle $, the
variance $V$ and the pair correlation $g_{12}=\left\langle \Delta\hat{n} 
_{1}\Delta\hat{n}_{2}\right\rangle /\sqrt{\left\langle \hat{n}_{1} 
\right\rangle \left\langle \hat{n}_{2}\right\rangle }$ are related by
$V=1-2g_{12}\left\langle \hat{n}_{1}\right\rangle /SN$. Here the shot noise
level for fermions is given by $SN=\sum_{i}\left\langle (\Delta\hat{n} 
_{i})^{2}\right\rangle =\sum_{i}\left\langle \hat{n}_{i}\right\rangle
(1-\left\langle \hat{n}_{i}\right\rangle )$, which we point out is always
sub-Poissonian and is independent of the state of the fermion system. For a
nonzero $SN$, the assumption of a perfect noise reduction below the shot-noise
level, $V=0$, can be used to identify the maximum degree of correlation,
giving $g_{12}^{(\max)}=1-\left\langle \hat{n}_{1}\right\rangle $. This is
exactly as obtained in the present model, Eq.~(\ref{g-diss}). At times when
the average occupancies approach one, the shot-noise itself vanishes.
Therefore, the notions of \emph{sub} shot-noise fluctuations and maximal
correlation become meaningless for $\left\langle \hat{n}_{i}\right\rangle =1$.

For comparison, in the bosonic case the shot-noise level corresponds to that
of a coherent state, $SN=\left\langle \hat{n}_{1}\right\rangle +\left\langle
\hat{n}_{2}\right\rangle $, and never vanishes. For equal mode occupancies, it
gives $V=1+\left\langle \hat{n}_{1}\right\rangle -g_{12}$ and therefore $V=0$
implies that $g_{12}^{(\max)}=1+\left\langle \hat{n}_{1}\right\rangle $. This
is always larger than the uncorrelated level of $0$, and again agrees with the
actual solution to the problem of molecule dissociation.

In order to apply the results of the present uniform model to realistic
trapped condensates, we remark that the quantization length $L$ should be
matched to the characteristic size of the molecular BEC. In addition, one has
to ensure that the time window for dissociation is chosen such that the
momentum kick $k_{0}\simeq\sqrt{2m|\Delta|/\hbar}$ imparted on the atoms is
not too large, so that the atoms created mostly near the trap center remain
within the molecular BEC while the dissociation is on. This implies that the
present results are applicable to $t\lesssim t_{\max}=Lm/(2\hbar k_{0})$ or
$\tau\lesssim\tau_{\max}=l/(4\sqrt{|\delta|})$. Considering a typical set of
parameters, with $3\times10^{5}$ initial number of molecules and $l\simeq46$
\cite{Parameters}, the example presented here for $\delta=-16$ would produce
$\sim1.5\times10^{4}$ atoms in each spin state at $\tau=0.6$ which compares
favorably with $\tau_{\max}\simeq3$. This corresponds to $5\% $ conversion and
is consistent with the use of the undepleted molecular field approximation.

In practice, the main factors which may contribute to the reduction of the
correlation signal are: (\textit{i}) the presence of a large thermal component
in the initial molecular gas, in which case the thermal centre-of-mass momenta
may no longer be negligible compared to the momentum kick of the atoms $k_{0} 
$; and (\textit{ii}) atom-atom $s$-wave scattering between the two spin
components, which at high densities may substantially redistribute the momenta
over the $s$-wave scattering spheres and spoil the correlations.

In summary, we have analyzed short time dynamics of dissociation of a BEC of
molecular dimers into correlated fermionic atoms in two different spin states.
The pair correlations between atoms with opposite spins and momenta calculated
here correspond to the maximum possible degree of correlation and serve as
upper bounds for more detailed calculations and comparisons with experiments.
The system may find applications in precision measurements beyond the
shot-noise level, as well as for fundamental tests of quantum mechanics with
macroscopic number of fermions, such as demonstrations of Bohm's version of
the Einstein-Podolsky-Rosen paradox \cite{EPR} and tests of Bell's
inequalities for spin observables.

The author acknowledges stimulating discussions with J. Corney, P. Drummond,
M. Greiner, and M. Olsen, and the support by the Australian Research Council.


\begin{thebibliography}{99}                                                                                               %


\bibitem {Greiner}M. Greiner \textit{et al.}, Phys. Rev. Lett. \textbf{94},
110401 (2005).

\bibitem {Bloch}S. F\"{o}lling \textit{et al.}, Nature (London) \textbf{434},
481 (2005).

\bibitem {Raizen}C.-S. Chuu \textit{et al.}, Phys. Rev. Lett. \textbf{95},
260403 (2005).

\bibitem {Esslinger}A. \"{O}ttl \textit{et al.}, Phys. Rev. Lett. \textbf{95},
090404 (2005).

\bibitem {Aspect}M. Schellekens \textit{et al.}, Science \textbf{310},
648 (2005).

\bibitem {HBT}R. Hanbury Brown and R. Q. Twiss, Nature (London) \textbf{177},
27 (1956).

\bibitem {Yasuda-Shimizu}M. Yasuda and F. Shimizu, Phys. Rev. Lett.
\textbf{77}, 3090 (1996).

\bibitem {Phillips-1997}E. A. Burt \textit{et al.}, Phys. Rev. Lett.
\textbf{79}, 337 (1997).

\bibitem {Kasevich}C. Orzel \textit{et al.}, Science \textbf{291}, 2386 (2001).

\bibitem {Tolra}B. L. Tolra \textit{et al.}, Phys. Rev. Lett. \textbf{92},
190401 (2004).

\bibitem {Weiss}T. Kinoshita, T. Wenger, and D. S. Weiss, Phys. Rev. Lett.
\textbf{95}, 190406 (2005).

\bibitem {Kimble}L.-A. Wu \textit{et al.}, Phys. Rev. Lett. \textbf{57}, 2520 (1986).

\bibitem {Heidman}A. Heidmann \textit{et al.}, Phys. Rev. Lett. \textbf{59},
2555 (1987).

\bibitem {JOSA-B-1987}Special issue on Squeezed States of the Electromagnetic
Field. \textit{J. Opt. Soc. Am. B} \textbf{4}, No. 10 (1987).

\bibitem {Walls}D. F. Walls and G. J. Milburn, \textit{Quantum Optics}
(Springer, Berlin, 1994).

\bibitem {Moelmer2001}U. V. Poulsen and K. M\o lmer, Phys. Rev. A \textbf{63},
023604 (2001).

\bibitem {TwinBeams}K. V. Kheruntsyan and P. D. Drummond, Phys. Rev. A
\textbf{66}, 031602(R) (2002); K. V. Kheruntsyan, Phys. Rev. A \textbf{71},
053609 (2005).

\bibitem {Yurovsky}V. A. Yurovsky and A. Ben-Reuven, Phys. Rev. A \textbf{67},
043611 (2003).

\bibitem {EPR}K. V. Kheruntsyan, M. K. Olsen, and P. D. Drummond, Phys. Rev.
Lett. \textbf{95}, 150405 (2005).

\bibitem {Dissociation-exp-Ketterle}T. Mukaiyama \textit{et al.}, Phys. Rev.
Lett. \textbf{92}, 180402 (2004).

\bibitem {Dissociation-exp-Rempe}S. D\"{u}rr, S. T. Volz, and G. Rempe, Phys.
Rev. A \textbf{70}, 031601(R) (2004).

\bibitem {FWM-fermions}M. G. Moore and P. Meystre, Phys. Rev. Lett.
\textbf{86}, 4199 (2001); W. Ketterle and S. Inouye, \textit{ibid.
}\textbf{86}, 4203 (2001).

\bibitem {Javanainen}J. Javanainen \textit{et al.}, Phys. Rev. Lett.
\textbf{92}, 200402 (2004).

\bibitem {Meystre-association}D. Meiser and P. Meystre, Phys. Rev. Lett.
\textbf{94}, 093001 (2005).

\bibitem {Jack-Pu}M. W. Jack and H. Pu, Phys. Rev. A \textbf{72}, 063625 (2006).

\bibitem {PDKKHH}P. D. Drummond, K. V. Kheruntsyan, and H. He, J. Opt. B:
Quantum Semiclass. Opt. \textbf{1}, 387 (1999); K. V. Kheruntsyan and P. D.
Drummond, Phys. Rev. A \textbf{61}, 063816 (2000).

\bibitem {Comment}These solutions refer to atom pairs in two different
(integer) spin states and are similar to those discussed in Refs.
\cite{TwinBeams,Yurovsky} for the case of the same spin state.

\bibitem {Corney-Drummond}J. F. Corney and P. D. Drummond, Phys. Rev. Lett.
\textbf{93}, 260401 (2004).

\bibitem {Parameters}These parameters can be obtained using a molecular BEC at
a peak density $n_{0}=5\times10^{19}$ m$^{-3}$ in an isotropic harmonic trap
of frequency $\omega/2\pi=70$ Hz, with a molecule-molecule $s$-wave scattering
length of $\sim57.4$ nm. In addition, we take $\chi\simeq2.5\times10^{-7}$
m$^{3/2}$/s, so that $g\simeq1.8\times10^{3}$ s$^{-1}$ and $t_{0}\simeq0.55$
ms. As a result, $\tau=0.6$ corresponds to $t=0.33$ ms duration of
dissociation, while $|\delta|=16$ converts to $|\Delta|/2\pi\simeq4.6$ kHz.
\end{thebibliography}
\end{document}